# Enhanced Surface Second Harmonic Generation in Nanolaminates

Fatemeh Abtahi,[1] Pallabi Paul,[1,2] Sebastian Beer,[1] Athira Kuppadakkath,[1] Anton Pakhomov,[1] Adriana Szeghalmi,[1,2] Stefan Nolte,[1,2] Frank Setzpfandt,[1,2] and Falk Eilenberger[1,2,3]

[1]Friedrich Schiller University Jena, Institute of Applied Physics, Abbe Center of Photonics, Albert-Einstein-Str. 15, 07745 Jena, Germany

[2]Franhofer-Institute for Applied Optics and Precision Engineering IOF, Albert-Einstein-Straße7, 07745 Jena, Germany

[3]Max Plank School of Photonics, Jena, Germany

*Supporting Information Placeholder*

**ABSTRACT:** Second-harmonic generation (SHG) is a second-order nonlinear optical process that is not allowed in media with inversion symmetry. However, due to the broken symmetry at the surface, surface SHG still occurs, but is generally small. We experimentally investigate the surface SHG in periodic stacks of alternating, subwavelength dielectric layers, which have a large number of surfaces, thus enhancing surface SHG considerably. To this end, multilayer stacks of $SiO_2/TiO_2$ were grown by Plasma Enhanced Atomic Layer Deposition (PEALD) on fused silica substrates. With this technique individual layers of a thickness of less than 2 nm can be fabricated. We experimentally show that under large angles of incidence (> 20 degrees) there is substantial SHG, well beyond the level, which can be observed from simple interfaces. We perform this experiment for samples with different periods and thickness of $SiO_2/TiO_2$ and our results are in agreement with theoretical calculations.

**KEYWORDS:** Surface Second Harmonic Generation, Ultrathin dielectrics, Plasma Enhanced Layer Deposition,

Second-harmonic generation (SHG) is a second-order nonlinear optical process which can only occur in materials and structures that have broken inversion symmetry [1,2]. In this process two photons with energy $\hbar\omega$ (wavelength $= \lambda = \frac{2\pi c}{\omega}$) will create one photon with energy $2\hbar\omega$ (wavelength $= \lambda/2$ ). SHG is forbidden in bulk materials with inversion symmetry [1,3,4], such as amorphous dielectric materials, like $TiO_2$ and $SiO_2$, unless some asymmetric is induced, e.g. by doping [5,6], poling [7,8], dc electric-fields [9,10], induced stress [11], etc.

Surfaces, however, induce structural discontinuity, leading to a local inversion symmetry breaking in the direction normal to the surface [3,12–19]. Hence surface-induced second harmonic has been observed in many centro-symmetric materials and structures such as dielectric metasurfaces [20], centrosymmetric particles in bulk isotropic solution (polystyrene) [21], metal-dielectric interfaces [22], ABC-type nanolaminates [23], Si-$SiO_2$ interfaces and many others. Moreover, it can be used to map specific properties of surfaces, such as the density and nature of dangling bonds [19]. Nevertheless, this effect is in general weak and hence applications in nonlinear optics have been elusive, so far.

A theoretical study conducted by Pakhomov et al, [24], which in turn is based on the study of surface Second Harmonic (SH) in layers systems [25,26] has proposed to enhance surface SH by functionalizing the surface of a dielectric substrate with a nanolaminate, hereby increasing the number of interfaces. A nanolaminate is a strongly subwavelength layer stack, which is so thin that interference effects are reduced, such that their linear behavior is more akin to an artificial material with averaged properties. One of their key findings was, that an enhancement of surface SH can be observed even for strictly periodic stacks of only two alternating materials, because adjacent surfaces experience non-negligible phase shifts and hence their contribution to the nonlinear polarization fields do not cancel exactly.

Here we experimentally study the enhancement of the surface SH of a glass substrate functionalized by nanolaminates, composed of periodic nanometer-scale layers of alternating $SiO_2$ and $TiO_2$ of different film thicknesses and numbers of periods. We have shown that despite AB symmetry in the structure, the SH signal of a such a nanolaminate is significantly enhanced. We find that the surface SHG of adjacent interfaces do not cancel, due to the small, yet finite phase shifts the light acquires between them. The increase in SH signal is in accordance with numerical models. Also, we show that the SH signal is highly dependent on the polarization of light, and the angle of incidence of the excitation beam, on the specific geometry of the nanolaminate.

## Sample Fabrication and Modeling

The nanolaminates were fabricated with the help of atomic layer deposition (ALD), with a thicknesses ranging from just above 40 nm down to 2 nm [27]. ALD is based on sequential self-limited surface reactions and can be used to grow high quality optical films [28] down to single nanometer thickness [29–31].

In our work we investigate four nanolaminates in two groups of samples with 5 and 10 layer pairs each. The thickness of $SiO_2$ is fixed at 2 nm for the 10 layer pair group and 5 nm for the 5 layer pair group. Within each group the thickness of $TiO_2$ is varied. The film thickness on a reference Si wafer was determined by spectroscopic ellipsometry (M2000, J.A. Woollam Co. Inc., Lincoln, NE, US). The obtained thickness is in a very reasonable agreement with the expected thickness. A table of the specific values is given in Table. 1 and a STEM-cross section through one sample of each group is displayed in Figure 1.

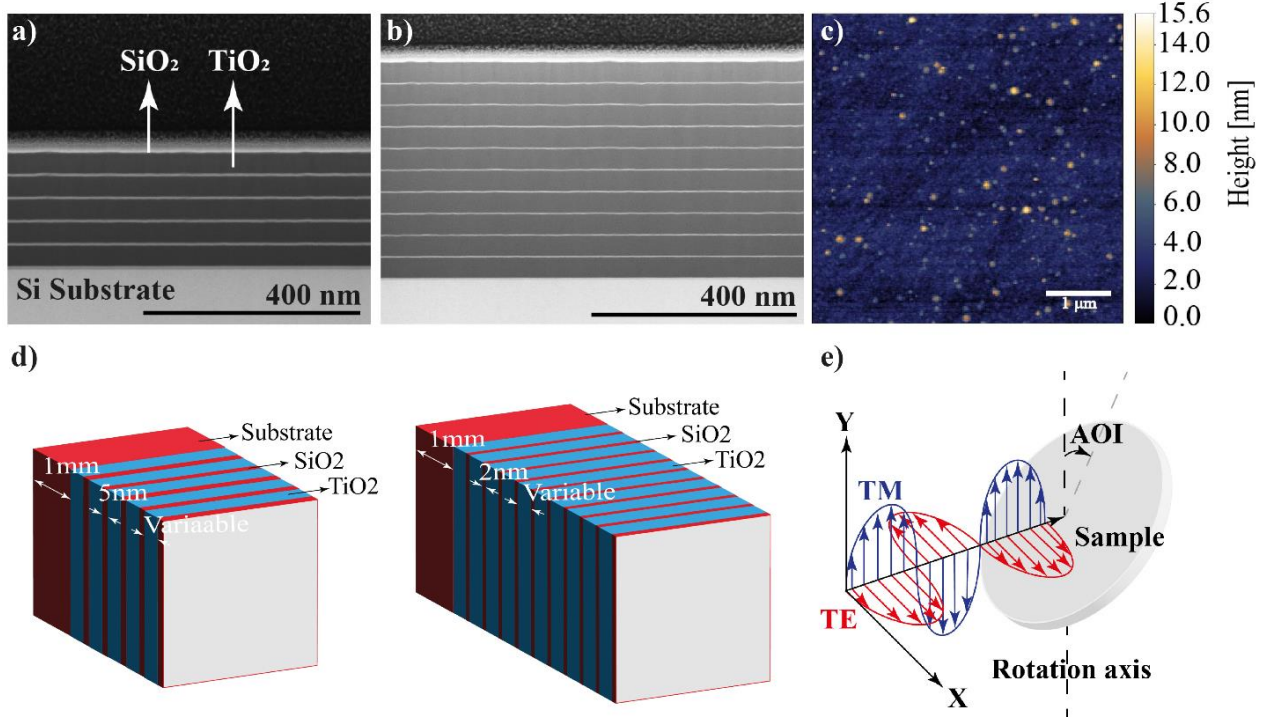

Table 1. Specifications of samples.

| Sample | SiO$_2$ [nm] | TiO$_2$ [nm] | Number of pair layers | Total target thickness [nm] | Thickness by ellipsometry [nm] |
|---|---|---|---|---|---|
| S01 | 2 | 31 | 10 | 330 | 345.6 |
| S02 | 2 | 25 | 10 | 270 | 283.1 |
| S03 | 5 | 33 | 5 | 190 | 196.7 |
| S04 | 5 | 17 | 5 | 110 | 113.5 |
| S00 | Substrate - Fused Silica- 1mm | | | | |

Figure 1. STEM-cross section graph of one sample of each group of samples. a) 5 pair layers of 5 nm SiO$_2$ and 33 nm TiO2 (S03) and b) 10 pair layers of 2 nm SiO$_2$ and 31 nm (S01), c) AFM map of the surface roughness of the top layer-SiO$_2$ for sample S01 with r.m.s =1.1 nm d) schematic of the samples for each group which shows the thickness of the substrate (fused silica) and other layers. e) TE and TM mode based on the rotation axis.

After fabrication the surface of the samples were analyzed with an AFM to determine their specific roughness and morphologic features. Overall, the surface quality was found to be good, with measured r.m.s roughness values on the range of 1.1 to 2 nm (Figure 1-c), which means that the samples have a high surface quality and the SHG process would not be affected by considerable amounts of scattering even for somewhat larger number of layer pairs as we demonstrated in this work (i.e. 10 pairs).

Specific values of the thickness for TiO$_2$ have been chosen from simulations of the SHG-efficiency based on the model from [24], where the thickness of TiO$_2$ and the angle of incidence was systematically varied. Results of the simulations for TM-polarization are displayed in Figure 2. The specific sample parameters, which have been chosen for fabrication are marked in the figure and are selected so as to coincide with characteristic maxima and minima of the surface SHG-efficiency.

The simulation is based on the below formula for the surface nonlinear polarization [24]:

$$\vec{P}_S^{NL}(2\omega,\vec{r}) = \varepsilon_0 \chi_{\perp\perp\perp}^{(2)} E_\perp(\omega,\vec{r})E_\perp(\omega,\vec{r}) \times \vec{n} \\ + \varepsilon_0 \chi_{\perp\parallel\parallel}^{(2)} E_\parallel(\omega,\vec{r})E_\parallel(\omega,\vec{r}) \times \vec{n} \\ + \varepsilon_0 \chi_{\parallel\perp\parallel}^{(2)} E_\perp(\omega,\vec{r})E_\parallel(\omega,\vec{r}) \times \vec{\tau}$$

where $\vec{n}$ and $\vec{\tau}$ are the outward normal unit vector and the tangent unit vector at the point $\vec{r}$ on the surface pointing in the direction of $E_\parallel(\omega,\vec{r})$, respectively.

In general, it can be observed that SHG efficiency is zero for normal incidence and then increases for larger angles of incidence, before eventually dropping off at grazing incidence. The former is because the nonlinear tensor has exclusively out-of-plane components, which cannot be excited for normal incidence, where all polarization vectors are in plane. Likewise, the model only predicts SHG for TM-waves, as only those have out-of-plane-polarization. The drop-off at larger angles of incidence is attributed to the strong Fresnel reflection at the first surface at such angles which prevents the light from penetrating into the nanolaminate and hence reduces SHG as well.

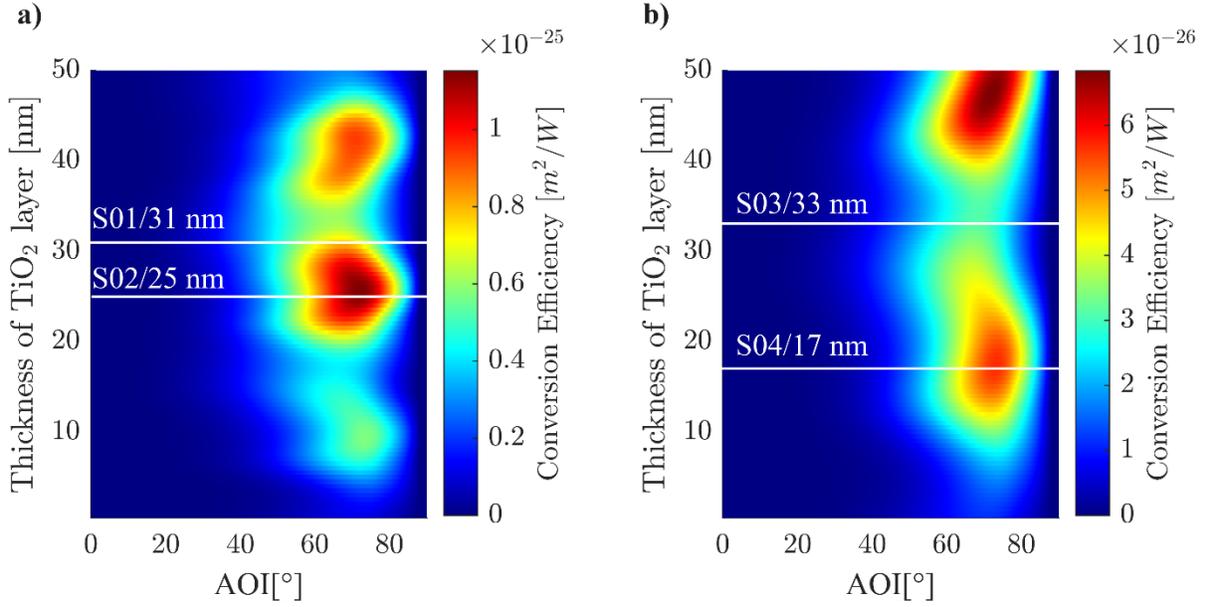

Figure 2. Heat map of the surface SHG for different thickness of the TiO$_2$, which is also varying depends on the different AOI, where the 4 specific samples in this work are marked a) 2 nm SiO$_2$ 10 pair layers and b) 5 nm SiO$_2$ with 5 pair layers.

**Experimental Investigation**

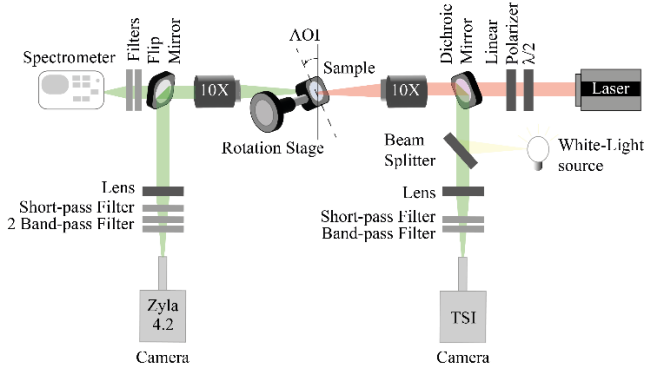

Figure 3. Experimental setup for nonlinear SHG microscopy of surface SHG in nanolaminates.

We characterized the angle-dependence of the surface SHG of the samples using a nonlinear microscopy setup, which is depicted in Figure 3 We excited the sample using a Ti-Sapphire laser with a pulse duration of T = 100 fs and up to P = 380 mW average power at a wavelength of $\lambda$ = 820 nm with a repetition rate of f = 80 MHz. The incident power and the polarization (TE or TM) was fixed by a combination of half wave plates and a polarizer. To focus the laser on the sample we are using a low magnification (10X), small NA (0.26) objective. The incidence angle can be tuned by rotating the sample with respect to the fixed incident beam from AOI = 0° (normal incidence) to AOI = 80° (grazing incidence). Special care is taken as to make sure that laser focus, sample surface and axis of rotation coincide. This is enabled by the installation of a wide-field imaging path with the help of a dichroic mirror, which can also be used to measure reflected SHG for very small angles of incidence. We measured the lasers spot diameter ($d_{laser\ beam}$ = 9.3 µm) to have an area of A = 68×10$^{-12}$ m$^2$ at an angle of incidence of AOI = 0°, which scales as $\sim 1/\cos(\text{AOI})$, when the angle of incidence is changed. At AOI = 0° we thus get a peak power density of P = 4.6× 10$^{14}$ W/m$^2$.

Surface SHG is collected in transmission using an identical objective which images onto a camera. We use a Zyla 4.2 Andor. The large field of view also accommodates for lateral beam walk during sample rotation due to refraction in the substrate. To block the fundamental signal from the incoming beam, we have used 3 filters, one short pass filter (cut off at 600 nm) and 2 band pass filters (380-610 nm).

Alternative to the detection with the camera we can also measure the SH signal with a spectrometer (FLAME Miniature Spectrometer, FLAME-T-VIS-NIR-ES, Ocean Insight), which is used to make sure we observe SHG at the appropriate wavelength. A measured spectrum is displayed in Figure 4-a, which clearly shows a residual peak at 820 nm and a strong SHG signal at 410 nm. Note that the residual Fundamental Wavelength (FW) peak is only visible in this measurement as a reference because we removed one bandpass filter from the setup. In a next step we conduct a power scaling measurement to further verify that we indeed observe SHG. Results are displayed in Figure 4-b together with a polynomial fit, which gives a slope of 2.01 ± 0.1, as expected for SHG.

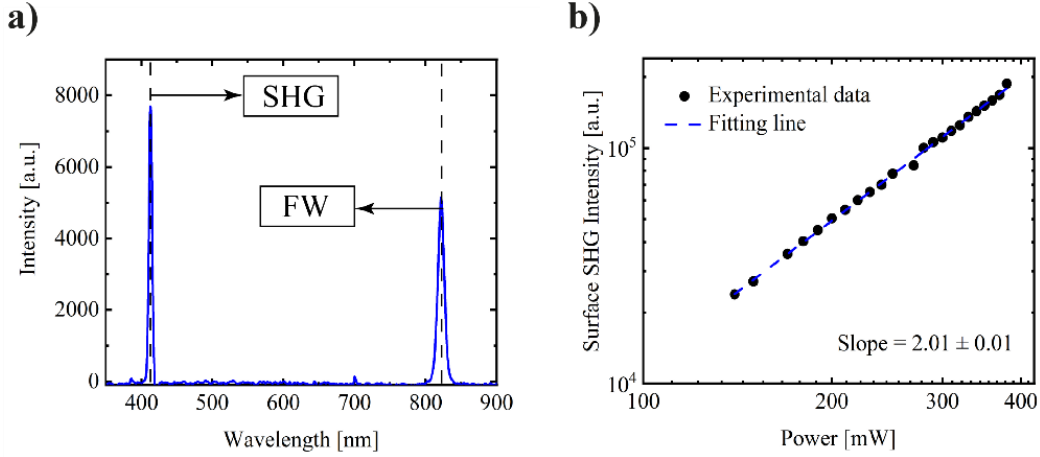

Figure 4. a) Spectrum of the fundamental and SH signal (with Spectrometer) and b) Power scaling measurement (with camera) for S04 (5 nm $SiO_2$, 17 nm $TiO_2$ with 5 layer pairs).

After accounting for the quantum efficiency of the camera, as well as the transmission efficiency of the optical system we proceed to determine the surface SH efficiency from the experimental data ('Supporting Information - Section 4'), which is given by:

$$\eta = \frac{P_{SH}}{P_{FW}^2} \times \frac{A_{FW}^2}{A_{SH}} \times \cos(AOI) \times T \times f \times 1.06\sqrt{2}$$

where all power values are average powers. $A_{SH}$ and $A_{FW}$ are the SH and FW focal spot area. The experimental results for two samples, measured in TM and TE polarization are displayed in Figure 5, together with simulated data. Keep in mind that the simulation required the input of a specific value $\chi^{(2)}$ for each interface, which we calculated by defining a fitting parameter with the data obtained for sample S03 in TM polarization and then reuse for all other samples under investigation. The specific value we obtain is $\chi^{(2)}_{\perp\perp\perp} = 4.20\times10^{-21}$ $m^2V^{-1}$ at 820 nm, which is representative for all $SiO_2/TiO_2$-interfaces in our experiment. The calculated values for other interfaces are placed in 'Supporting Information – Section 3'.

The efficiency for the surface SH is dramatically different for the TM- and TE-case, as expected. While there is significant surface SHG for TM-polarization, where a part of the polarization vector is perpendicular to the surface and hence probes the discontinuity, this is not the case for TE polarization. Residual TE SHG is attributed to surface roughness and imperfect settings of the input polarization. Moreover, we see that experimental and numerical data agree very well with respect to the angular behavior. While there is no surface SHG at normal incidence this grows for larger AOIs until a drop is observed in the range of AOI ≈ 70°, where Fresnel reflection starts to deflect the FW away from the surface before SH can be generated.

Data is given for two representative (S02 with a 2 nm $SiO_2$ layer and 10 layer pairs and S04 with 5 nm $SiO_2$ layers and 5 layer pairs), the geometries of which are displayed in the inset of the respective graphs. It can be noted that the conversion efficiency is roughly comparable, as was predicted by the numerical model.

We proceed to measure all samples to confirm the validity of the model for a larger class of samples. The results are displayed in Figure 6-a, where the data from the numerical model is plotted in solid lines and the experimental data is plotted in dots. It is apparent that the numerical and experimental data coincide well, given the uncertainty of fabrication process and the specific properties of the material interfaces determined by this process.

As was already discussed, we observe a very good similarity between numerical and experimental data. However, as can already been seen from the complex features of the simulated graphs in Figure 2, there is no simple dependency on either the number of interface nor on the specific thickness of the $SiO_2$ or $TiO_2$ layers. This is attributed to the fact that our samples have a thickness range of roughly 110 to 350 nm with a corresponding optical thickness in the range of 140 nm to just below 1 μm. This is at the upper limit of the nanolaminate definition hence, particularly for the thicker nanolaminates, linear interference of the FW and of the SH plays an important role. Since we, however, have shown that the model we used is suitable to predict the conversion efficiency of our nanolaminates, we argue that in a next step we could proceed to even thinner samples, which more strictly adhere to the definition of nanolaminates or one could utilize the complex interference behavior of such mesoscopic structures to tailor specific resonances on the angular or wavelength domain.

To have a better perspective we can compare the data at a specific angle, which from our results we can see that they are in a good comparison with the simulation (Figure 6-b). Moreover, this graph also shows amount of SHG signal produced by a blank fused silica substrate, which is more than 10 times weaker than the signal we observe in sample S03. Hence, we claim that dielectric nanolaminates can be used to enhance the surface SHG by up to an order of magnitude.

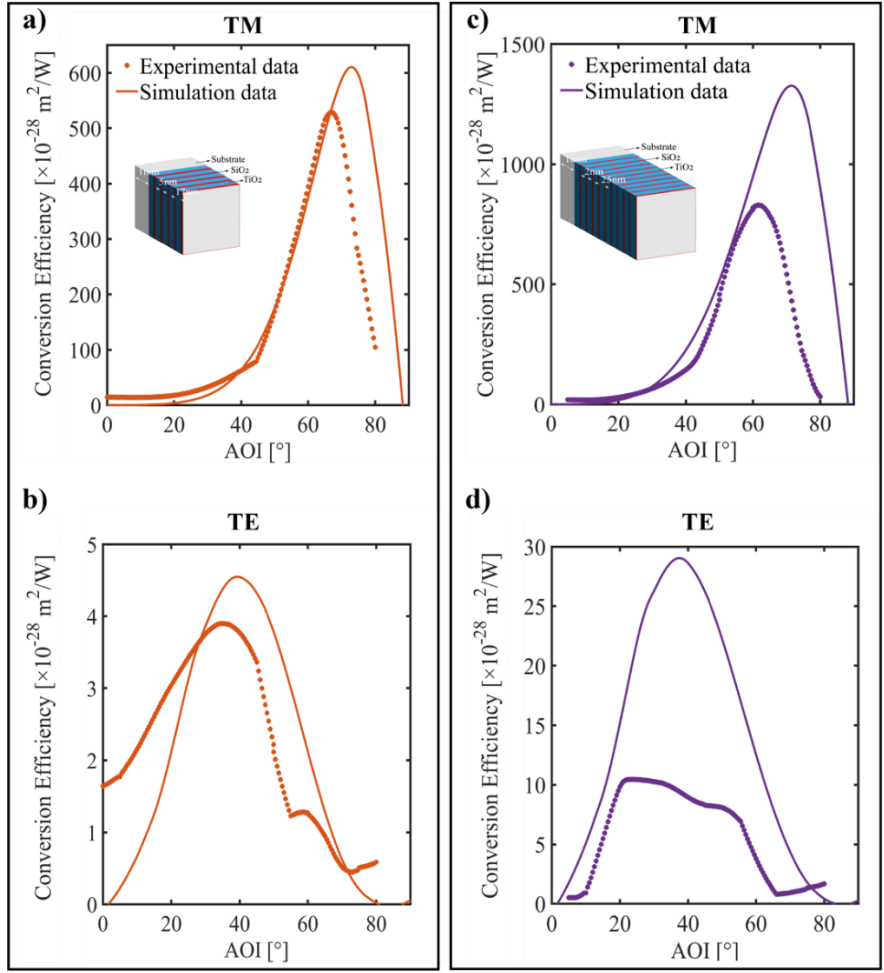

Figure 5. Experimental (dots) and simulation (solid line) data. a) TM Mode of sample 04 (5 nm $SiO_2$ –17 nm $TiO_2$) ×5 b) TE Mode of sample 04. c) TM Mode of sample 02 (2 nm $SiO_2$ – 31 nm $TiO_2$) ×10. d) TE Mode of sample 02.

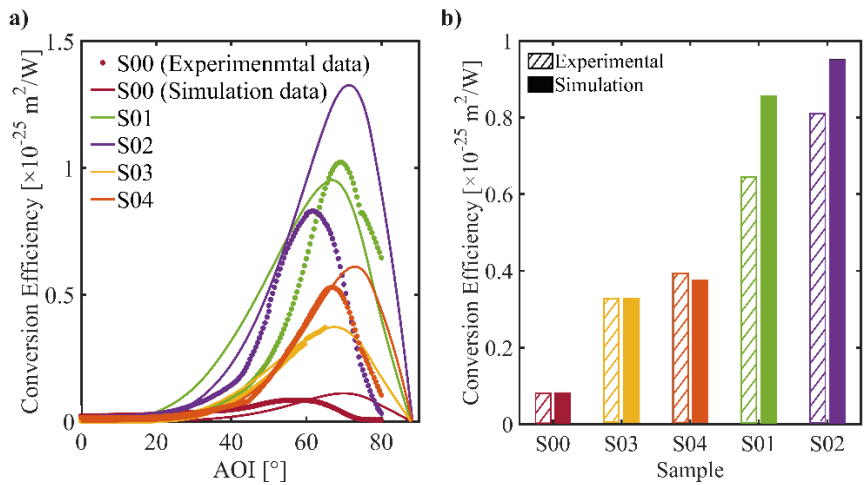

Figure 6. The results for all other samples are summarized here for 820 nm b) Comparing Simulation and Experimental data at AOI = 60° for all samples

## Conclusion

We have measured the surface Second Harmonic (SH) from binary dielectric nanolaminates, i.e., structures made of only 2 alternative amorphous materials, of subwavelength thickness. We have shown that, in accordance with numerical models, the SHG signal of such a nanolaminate is substantially enhanced, although the structure only has an AB symmetry. We find that the surface SHG of adjacent interfaces do not cancel, because of the small, yet finite phase shift the light acquires between them. We show that the SH signal is highly dependent on the angle of incidence of the excitation beam, on the specific geometry of the nanolaminate, and the polarization of light. Moreover, we can use the signal to fix the value of the out-of-plane component of the SHG-tensor of an $SiO_2/TiO_2$ interface to a value of $4.20\times10^{-21}$ $m^2V^{-1}$.

## AUTHOR INFORMATION

### Corresponding Author

**Fatemeh Abtahi** - Friedrich Schiller University Jena, Institute of Applied Physics, Albert-Einstein-Str. 15, 07745 Jena, Germany;

Email: Fatemeh.alsadat.abtahi@uni-jena.de

**Falk Eilenberger** - Friedrich Schiller University Jena, Institute of Applied Physics, Albert-Einstein-Str. 15, 07745 Jena, Germany;

Email: Falk.eilenberger@uni-jena.de

Notes
**The authors declare no competing financial interests.**

Funding. **Funded by the Deutsche Forschungsgemeinschaft (DFG, German Research Foundation) – Project-ID 398816777 – SFB 1375 and the Fraunhofer Society (FhG, Attract 066-601020)**

**Acknowledgments.** Thanks to Michael Steinert for the STEM investigation.

**Supporting Information.**